# ORBIT AND DYNAMIC ORIGIN OF THE RECENTLY RECOVERED ANNAMA'S H5 CHONDRITE

Josep M. Trigo-Rodríguez[1]
Esko Lyytinen[2]
Maria Gritsevich[2,3,4,5,8]
Manuel Moreno-Ibáñez[1]
William F. Bottke[6]
Iwan Williams[7]
Valery Lupovka[8]
Vasily Dmitriev[8]
Tomas Kohout[2,9,10]
Victor Grokhovsky[4]

[1] Institute of Space Science (CSIC-IEEC), Campus UAB, Facultat de Ciències, Torre C5-parell-2ª, 08193 Bellaterra, Barcelona, Spain. E-mail: trigo@ice.csic.es
[2] Finnish Fireball Network, Helsinki, Finland.
[3] Dept. of Geodesy and Geodynamics, Finnish Geospatial Research Institute (FGI), National Land Survey of Finland, Geodeentinrinne 2, FI-02431 Masala, Finland
[4] Dept. of Physical Methods and Devices for Quality Control, Institute of Physics and Technology, Ural Federal University, Mira street 19, 620002 Ekaterinburg, Russia.
[5] Russian Academy of Sciences, Dorodnicyn Computing Centre, Dept. of Computational Physics, Valilova 40, 119333 Moscow, Russia.
[6] Southwest Research Institute, 1050 Walnut St., Suite 300, Boulder, CO 80302, USA.
[7] Astronomy Unit, Queen Mary, University of London, Mile End Rd. London E1 4NS, UK.
[8] Moscow State University of Geodesy and Cartography (MIIGAiK), Extraterrestrial Laboratory, Moscow, Russian Federation
[9] Department of Physics, University of Helsinki, P.O. Box 64, 00014 Helsinki University, Finland
[10] Institute of Geology, The Czech Academy of Sciences, Rozvojová 269, 16500 Prague 6, Czech Republic

**Abstract:** We describe the fall of Annama meteorite occurred in the remote Kola Peninsula (Russia) close to Finnish border on April 19, 2014 (local time). The fireball was instrumentally observed by the Finnish Fireball Network. From these observations the strewnfield was computed and two first meteorites were found only a few hundred meters from the predicted landing site on May 29$^{th}$ and May 30$^{th}$ 2014, so that the meteorite (an H4-5 chondrite) experienced only minimal terrestrial alteration. The accuracy of the observations allowed a precise geocentric radiant to be obtained, and the heliocentric orbit for the progenitor meteoroid to be calculated. Backward integrations of the orbits of selected near-Earth asteroids and the Annama meteoroid showed that they rapidly diverged so that the Annama meteorites are unlikely related to them. The only exception seems to be the recently discovered 2014UR116 that shows a plausible dynamic relationship. Instead, analysis of the heliocentric orbit of the meteoroid suggests that the delivery of Annama onto an Earth-crossing Apollo type orbit occurred via the 4:1 mean motion resonance with Jupiter or the nu6 secular resonance, dynamic mechanisms that are responsible for delivering to Earth most meteorites studied so far.

Introduction.

The recovery of a meteorite following an accurate trajectory reconstruction of its luminous bolide phase is rarely achieved. Recovery is even less frequent when the observations come from a continuous monitoring effort made by a ground-based fireball network. At the time of writing, 21 meteorites exist where the heliocentric orbit has been calculated from observations of the fireball generated by the passage of the meteoroid through the Earth's atmosphere and details of these are given in Table 1. In order for a meteorite to survive the passage through the Earth's atmosphere, the initial meteoroid must have been large, producing a very bright fireball or superbolides with a luminous magnitude over -16. These are rare and unpredictable both in time and location so that the accuracy and reliability of the observations vary widely from event to event. Some are relatively good while others are less trustworthy, depending on the type, quality and number of records. Some events were imaged by accident by untrained observers, but a significant number came about through programs that regularly monitor the skies. Modern digital cameras have allowed casual images to be obtained even in day-time, producing valuable records of the luminous fireball phase that might be calibrated (see e.g. Trigo-Rodríguez et al., 2006). Even though the fireballs are extremely bright, the progenitor meteoroids are still less than a few meters across, usually too small to be recognized by telescopic monitoring programs searching for potential threats to the Earth. However, there are some exceptions such as 2008 TC3 that was a small asteroid about 5-meters across that disrupted over the Nubbian desert and produced the Almahatta Sita meteorite.

From all the meteorites with known orbit listed in Table 1, excluding Annama, 8 of them are H type ordinary chondrites. Grady (2000) found that 31.4% of meteorite falls are H type chondrites, thus we should expect 6 or 7 to be represented. As we are dealing with statistics of small numbers the difference is not significant. It is plausible that all the H chondrites come from a progenitor asteroid that fragmented into several pieces. Indeed, there is spectral and geochemical evidence that the H chondrites and the IIE iron meteorites may originate from asteroid 6 Hebe (Gaffey and Gilbert, 1998). Progressive disruption and resonance effects could have scattered enough small asteroid fragments for being today one of the most common meteorite groups delivered to Earth. The petrologic type of this group of ordinary chondrites are additionally classified from 3 to 6, depending on the different degrees of thermal metamorphism.

Table 1.

The possibility that meteorite-dropping bolide complexes associated with asteroids could exist was first proposed by Halliday (1987). Trigo-Rodríguez et al. (2007, 2008) also found dynamic associations between large meteoroids and Near Earth Objects (NEOs). Many asteroids are rubble piles and so probably do not require a collision in order to be disrupted. The fragmentation process is likely to produce many meter-sized rocks as well as larger boulders and rubble pile asteroids that could form a complex of asteroidal fragments once disrupted all initially moving on nearly identical orbits. Detecting such families or associations may not be easy because the life time of such orbital complexes is quite short (few tens of thousand of years) as consequence of planetary perturbations (Pauls and Gladman, 2005), except perhaps for those cases exhibiting orbits with high inclination, where life-times can be considerably higher (Jones and Williams, 2008), while disruptive and collisional processes also cause a divergence in the orbits (Bottke et al., 2002). Significant brecciation, and shock-induced darkening has been found e.g. in Almahata Sitta or Chelyabinsk meteorites (Kohout et al., 2014; 2010; Bischoff et al., 2010; Horstmann and Bischoff, 2014) indicating that collisions played a role in their evolution.

Other mechanisms for delivering meter-size meteoroids to Earth include tidal fracturing caused by close encounters with planets and fast rotation (Trigo-Rodríguez et al., 2007). Catastrophic disruptions are characterized by most of the initial mass being ejected away at escape velocity (Bottke et al., 2005) which is considerably smaller than the orbital velocity. Consequently, metre-sized pebbles or larger boulders are released forming a stream of asteroidal fragments moving on nearly identical orbits (Williams 2002, 2004; Jenniskens, 2006).

In this paper we present trajectory and orbital data for Annama's progenitor meteoroid obtained from the fireball imagery (Gritsevich et al., 2014a). The fall of Annama meteorite occurred in the remote Kola Peninsula (Russia) close to Finnish border on April 19, 2014 (local time) and the fireball was observed by the Finnish Fireball Network as well as numerous local residents. From these observations the strewnfield was computed and two first meteorites were found only about a few hundred meters from the predicted landing site on May 29$^{th}$ and May 30$^{th}$ 2014. The meteorites were later characterized as H5 chondrite (Gritsevich et al. 2014a). To gain insight in the origin of H chondrites in the near-Earth vicinity we also explore the possible existence of Near Earth Asteroids (NEAs) capable of producing Annama's meteoroid. The description of the observational methods, reduction procedures, and results are given in next section. In section 3, the discussion and main implications of the results are given in the context of the sources of meteorites reaching the Earth. Finally, some general conclusions on this new meteorite are presented.

2. Observations and data reduction.

Continuous monitoring of the skies for meteor and fireball activity over Finland was initiated by Ilkka Yrjölä and has continued with different degrees of coverage from 1998. This became incorporated into the Finnish Fireball Network (hereafter, FN) in 2002. Today the network monitors a surface of about 400.000 km$^2$ with most of the observations made by amateur astronomers (Gritsevich et al. 2014b). The event under discussion, initially named the Kola Peninsula fireball (FN20140419), was imaged on April 18, 2014 at 22h14m09.3±0.1s UTC from three FN stations: Kuusamo, Muhos, and Mikkeli. It was fortunate that an additional dashcam recording was made by Alexandr Nesterov in Snezhnogorsk, Russia. The locations of these are given in Table 2, while an image from the dashcam is shown as Figure 1. The general camera details and resolution computed from the calibrations are given in Supplementary Table 2b.

After the initial registration of the fireball, the fb_entry program which has been validated using different types of observational data, including most of the fireball cases imaged in Finland (Lyytinen and Gritsevich, 2013) was used to analyze the observations. In the past, the program was applied to the number of cases, including observations which may not be accurately timed, some may be observations from one station only, while some of the observations only had directions in use. There are three basic functions in the fb_entry program. One determines the general direction of the trajectory, its location and the velocity of the meteor along the early part of its path. The other two functions determine the individual velocities of the larger surviving fragments, when applicable. From this basic data, other physical parameters are determined using the methods described in Gritsevich (2007, 2009). The case of Annama was very interesting since the key scaling parameters as well as the terminal height of the fireball derived from our analysis were found to be practically the same as the corresponding values earlier reported by Gritsevich (2008) and Moreno-Ibáñez et al. (2015) for Innisfree, the only meteorite successfully recovered by the MORP program (Halliday et al., 1978).

All the images were calibrated using background star field, the astrometric measurements being made manually. The model assumes a symmetry point in the image that could be offset from the image center. The radial distance from this point depends on the weighted average of the equidistant and gnomonic projections, which in many cases gives quite a good approximation. In the wide field cameras the weight of the equidistant part is typically more than one and correspondingly the weight of the gnomonic part is negative. This radial model is further improved by means of a polynomial fit with powers up to either 5 or 7 as required. The pixel X/Y-ratio is also derived and in video cameras is often not 1.0.

Good calibrations (with an accuracy of a few hundredths of a degree) were obtained for both the Kuusamo and the Mikkeli cameras. In the Muhos camera field of view there were few stars near the fireball direction. This did not affect the final solution, because it was used mainly for checking purposes. The calibration of the video from Snezhnogorsk was much more difficult because of the lack of stars in the images. Stacking of a few frames might have been possible but because Jupiter just barely could be detected it was no hope of getting stars from only a few stacked images. Fortunately, Jupiter was found in a few frames before the main fireball outburst took place. The main calibration was made for one frame. The car location and azimuth directions were derived by means of Google-satellite images and Yandex images. All calibration directions were assigned with the azimuth and elevation values. It is very helpful, if the image contains vertical lines that can be measured at more than one point if the image is scarce in star data and this was the case with several buildings providing such vertical lines as can be seen in figure 1b.

The actual horizon was not visible, but one point with elevation 0 degrees was derived by means of some perspective properties of assumed horizontal directions such as roof-tops and window edges. The elevation of Jupiter was known. Jupiter was at a very different azimuth direction compared to the fireball, which has some disadvantages but also could be advantageous since the fit of this to other directions is quite sensitive to possible car location error. There were in all 13 calibration points, including Jupiter and the zero-elevation point (Fig. 1). One of the azimuth directions had three measure points at different elevation angles and three others had two. The pixel X/Y ratio was assumed 1.00, but it was also tested to be a free variable and the resulting value was very close to this. For other nearby frames, the fireball direction was transformed to the calibration by means of some distant terrestrial reference points close to the direction where the car was moving. One of the fireball directions was shifted to this frame by means of cloud patterns very near the fireball. The video frames were measured from 1600×900 pixel size images. From this calibration, fireball directions for four different video frames were measured. The RMS error in the azimuth calibration was $0.14^o$, the largest being $0.34^o$. The total azimuth directions span was more than 70 degrees, also covering the measured fireball azimuth range. The scale in the image center is 17.3 pixels/degree (see Table 2b). However, the highest apparent error of this magnitude in the calibration is due to the uncertainty in the actual directions measurements (from Google images) and not associated with the astrometric video accuracy.

The very first fireball direction was measured from a different car location and consequently the accuracy of this direction is not as good as the others and has been given a smaller weight in the trajectory derivation.

The fireball trajectory was computed using the fb_entry software developed by Lyytinen and Gritsevich (2013). In this case the entry track result is very similar to that obtained using the method of planes developed by Ceplecha (1987). Later on the orbital elements were computed using the Spanish Meteor Network software (Madiedo et al., 2011) as well as with the recently developed and tested software "Meteor Toolkit" (Dmitriev et al., 2014). The results are given in Tables 3 and 4.

3. Discussion.

When Drummond (1982) first calculated meteor radiants for bodies approaching within 0.2 AU of the Earth's orbit, the number of known asteroids was quite small, and only three orbits of meteorite-dropping bolides (Příbram, Lost City and Innisfree) were known as can be seen from the chronologic list shown in table 1. Fortunately, present achievements in completing the NEO inventory and size distribution (Bottke et al., 2002) together with a significant increase in fireball studies and meteorite recoveries provide dynamic clues on the origin of meter-sized meteoroids that might be source of localized hazard (Chapman, 2004, 2008; Brown et al., 2013).

3.1. Source regions for the Annama meteoroid.

To study the origin of Annama in the solar system we used an unpublished model based on the ideas developed by Morbideli and Gladman (1998). Assuming Annama's orbit was (a, e, i) = (1.990 AU, 0.690, 11.650°), we found that it was a 73% chance of coming from the nu6 resonance, and 27% chance of coming from the 3:1 resonance. On the other hand, we also looked at what would predict (Bottke et al. 2002) model even although it was meant for large bodies, not meteoroids, so results are given with caution here. The probabilities of having Annama coming from the JFC, outer main belt, 3:1 resonance, intermediate Mars-crossing region ($P_{MC}$), and the nu6 resonance ($P_{N6}$) were explored with such model. The model included a source called the "intermediate Mars-crossing region" that is probably not applicable to meter-sized rocks, so we just add this probability to the nu6 resonance for clarity. By adding the $P_{MC}$ and $P_{N6}$ together, it looks like the strongest probability is that the Annama meteoroid came from the innermost region of the main belt (i.e., it escaped near the nu6 resonance). The probabilities for the 3:1 resonance are relatively low: about 10-26% as a function of the uncertainty in Annama's orbital elements. Consequently, both results about the origin of Annama are consistent and temptatively suggest that from a source standpoint, and for H chondrites, one could argue that Annama potentially came from the same broad source region as Lost City, Peekskill, and Buzzard Coulee (Table 1). This result confirms the role of the nu6 secular resonance as source of meteorites envisioned long ago by Scholl and Froeschlé (1991). It is important to remark that Příbram, Moravka, and Grimsby seem to have a different source because their orbits are more likely associated with the 3:1 resonance.

3.2. Orbital clues on the origin of Annama: is there a link with known NEOs?

As can be seen in Fig. 2 the Annama meteoroid has an orbit that is quite similar to a typical Apollo NEO. For this reason we searched through the NEO databases for asteroid orbits that could be regarded as a present-day match for the derived Annama orbit, using the D-criterion of Southworth and Hawkins (1963). In reality, any of the criteria outlined in Jopek and Williams (2013) could be used, but the Southworth and Hawkins criterion has been used several times before in this context (e.g. Trigo-Rodríguez et al., 2007; Madiedo et al., 2013, 2014).

As was pointed out by Porubcan et al. (2004, 2006) before any association should be claimed, the orbital evolution should be similar for at least 5,000 years back in time. First of all we identified 12 potential candidates having D-criterion lower than 0.2 among the currently known NEOs: 2000EJ26, 2002EB3, 2002GM5, 2003GR22, 2004HA1, 2004VY14, 2005TU50, 2006JO, 2006WK130, 2012TT5, 2013LY28 and the recently discovered 2014UR116. The evolution of the orbits of these NEOs was calculated by numerical integrations using the Mercury 6 program (Chambers, 1999) a hybrid symplectic integrator widely used in Solar System dynamics studies. The orbits of the Annama bolide (including uncertainty in its pre-atmospheric velocity) and the orbits of the above listed NEOs were integrated back for at least 20,000 years. Perturbations from the planets Venus, Earth, Mars, Jupiter and Saturn were taken into account.

The D-criterion for all candidates was initially low at current time, but most exhibited a considerable divergence in few thousand years. However, that was not the case for the recently discovered PHA 2014UR116, where the D-criterion remained low throughout the integration, thus there may be a plausible connection with Annama (Fig. 3). For showing better the evolution this plot and the following one only goes backwards for 10,000 years that is enough for this overall discussion of the evolutionary trends. The results of these integrations for q, e, and i are shown in Fig. 4 for the range of pre-atmospheric velocity ($V_{inf}$) given by the uncertainty of Annama that we will call here "clones" for simplicity. The general evolution of the inclination is remarkably similar to 2014UR116 and seems to get closer for the Annama clones derived for the higher $V_{inf}$. This is what we should expect, since usually these measurements can underestimate the geocentric velocity at top of the atmosphere. From the inclination graph it seems that the best D-criterion match occurs for a Vinf=24.7 km/s. On the other hand the 24.6 km/s clone exhibits very different evolution. This clone matches very well the i, e, q values before diverging abruptly in inclination about 4500 years ago. Interestingly Fig. 3 shows that the lower $V_{inf}$ clones have a lower D-criterion values (D<0.3) over a short timescale of about 3,000 years. However the best short-term candidate is again the 24.6 km/s clone, but the D-criterion increases very quickly probably because of a close approach to one of the terrestrial planets.

4. Conclusions.

The pre-impact orbit derived for the Annama H5 chondrite and its backward analysis together with that of several NEAs give us the following clues on the origin of this meteoroid:

a) The Annama fireball was produced by a meteoroid with significant initial velocity (24.2 km/s) that came from an Apollo type orbit.
b) Backwards integration of the orbital elements of the progenitor meteoroid have identified that the PHA 2014UR116 could share a similar dynamic origin with Annama, but close approaches with terrestrial planets make it difficult to establish any other relationship among both bodies.

c) Trajectory reconstruction of Annama's bolide leaded to meteorite recovery of a meteorite that is the ninth H ordinary chondrite with accurate orbital elements. From the comparison with the eight previous H chondrites recovered we can conclude that Annama comes from the same broad source region as Lost City, Peekskill, and Buzzard Coulee, and it was delivered from a main belt resonance.
d) Considering Annama's orbital elements, a source probabilistic model suggests that the Annama meteoroid was delivered to Earth via the nu6 resonance with about a 70% of probability.

Acknowledgements

We thank the members of the Finnish Fireball Network who made this work possible through their efforts and also the members of the meteorite recovery expedition from Ekaterinburg. In particular we would like to thank Jarmo Moilanen, Steinar Midtskogen, Nikolai Kruglikov, Alexei Ischenko, Grigory Yakovlev, Jakub Haloda, Patricie Halodova, Jouni Peltoniemi, Asko Aikkila, Aki Taavitsainen, Jani Lauanne, Marko Pekkola, Pekka Kokko, Panu Lahtinen and Alexandr Nesterov. We are also grateful to José María Madiedo for his help with the asteroid backward integrations. This study is supported by the Spanish grant AYA 2011-26522, by the Academy of Finland project Nos 260027 and 257487, by the Finnish Geodetic Institute and by the Ministry of Education, Youth and Sports of the Czech Republic grant no. LH12079. Orbit determination using software "Meteor Toolkit" was carried out by Vasily Dmitriev, Valery Lupovka and Maria Gritsevich at MIIGAiK under the support of the Russian Science Foundation, project No. 14-22-00197 "Studies of Fundamental Geodetic Parameters and Topography of Planets and Satellites".

# TABLES

Table 1. Chronologic list of recovered meteorites with accurate orbital information. The uncertainty in each orbital element is not given here for simplicity, but it is implicit in the last figure given, and in the respective references. The name of other H chondrite falls appear in bold. Reference list: [1] Ceplecha (1961); [2] Spurný et al. (2003); [3] McCrosky et al. (1971); [4] Halliday et al. (1978); [5] Brown et al. (1994); [6] Hildebrand et al. (2006); [7] Borovicka et al. (2003); [8] Brown et al. (2004); [9] Trigo-Rodríguez et al. (2006); [10] Bland et al. (2009); [11] Jenniskens et al. (2009) and NEO JPL database for 2008 TC3; [12] Milley et al. (2010) [13] Haack et al. (2010); [14] ; [15]; [16] Spurný et al. (2011); [17] Borovička et al. (2013); [18] Jenniskens et al. (2012); [19] Jenniskens et al. (2014); [20] Borovička et al. 2013b.

| Meteorite Name | Year of Fall | Type | $V_g$ (km/s) | Orbital elements | | | | | | Reference |
|---|---|---|---|---|---|---|---|---|---|---|
| | | | | q (AU) | 1/a (AU$^{-1}$) | e | i (°) | ω (°) | Ω (°) | |
| Příbram | 1959 | H5 | 17.43 | 0.78951 | 0.416 | 0.6711 | 10.482 | 241.75 | 17.79147 | [1], [2] |
| Lost City | 1970 | H5 | 14.2 | 0.967 | 0.602 | 0.417 | 12.0 | 161.0 | 283.0 | [3] |
| Innisfree | 1977 | L5 | 14.2 | 0.986 | 0.534 | 0.4732 | 12.27 | 177.97 | 316.80 | [4] |
| Peekskill | 1992 | H6 | 14.7 | 0.886 | 0.671 | 0.41 | 4.9 | 308 | 17.030 | [5] |
| Tagish Lake | 2000 | C2-ung | 15.8 | 0.884 | 0.505 | 0.55 | 2.0 | 224.4 | 297.9 | [6] |
| Morávka | 2000 | H5 | 19.6 | 0.9823 | 0.541 | 0.47 | 32.2 | 203.5 | 46.258 | [7] |
| Neuschwanstein | 2000 | EL6 | 20.95 | 0.7929 | 0.417 | 0.670 | 11.41 | 241.20 | 16.82664 | [2] |
| Park Forest | 2003 | L5 | 16.1 | 0.811 | 0.395 | 0.680 | 3.2 | 237.5 | 6.1156 | [8] |
| Villalbeto de la Peña | 2004 | L6 | 16.9 | 0.860 | 0.435 | 0.63 | 0.0 | 132.3 | 283.6712 | [9] |
| Bunburra Rockhole | 2007 | Eucrite | 13.4 | 0.6428 | 1.175 | 0.245 | 9.07 | 209.87 | 297.59528 | [10] |
| Almahata Sitta | 2008 | Ureilite-an | 12.42 | 0.8999 | 0.7644 | 0.31206 | 2.5422 | 234.448 | 194.10114 | [11] |
| Buzzard Coulee | 2008 | H4 | 18.0 | 0.961 | 0.8130 | 0.22 | 25.5 | 212.0 | 238.9 | [12] |
| Maribo | 2009 | CM2 | 28.5 | 0.481 | 0.45 | 0.8 | 0.26 | 99.0 | 117.64 | [13] |
| Grimsby | 2009 | H5 | 20.9 | 0.9817 | 0.490 | 0.518 | 28.07 | 159.865 | 182.9561 | [14] |
| Jesenice | 2009 | L6 | 13.78 | 0.9965 | 0.571 | 0.431 | 9.6 | 190.5 | 19.196 | [15] |
| Mason Gully | 2010 | H5 | 14.53 | 0.98240 | 0.405 | 0.6023 | 0.832 | 18.95 | 203.2112 | [16] |
| Košice | 2010 | H5 | 10.3 | 0.957 | 0.369 | 0.647 | 2.0 | 204.2 | 340.072 | [17] |
| Sutter's Mill | 2012 | C | 28.6 | 0.456 | 0.386 | 0.824 | 2.38 | 77.8 | 32.774 | [18] |
| Novato | 2012 | L6 | 13.67 | 0.9880 | 0.478 | 0.526 | 5.51 | 347.35 | 24.9900 | [19] |
| Chelyabinsk | 2013 | LL5 | 19.03 | 0.738 | 0.581 | 0.571 | 4.98 | 107.67 | 326.459 | [20] |
| *Annama* | 2014 | H5 | 24.2 | 0.634 | 0.503 | 0.69 | 14.7 | 264.8 | 28.611 | This work |

Table 2.- Locations of the FN stations and the Russian dashcam which collected the videotapes analyzed here.

| Station # | Station (Province, country) | Camera operator | Longitude | Latitude (N) | Altitude (m) |
|---|---|---|---|---|---|
| 1 | Flotskaya str., 184682 Snezhnogorsk, Russia | Alexandr Nesterov | 33.24140 | 69.19484 | 80 |
| 2 | Kuusamo, FN, Finland | Asko Aikkila | 29.71819 | 65.94764 | 256 |
| 3 | Mikkeli, FN, Finland | Aki Taavitsainen, Jani Lauanne | 27.23953 | 61.68440 | 148 |
| 4 | Muhos, FN, Finland | Pekka Kokko | 26.01337 | 64.95493 | 71 |

Table 2b. Camera data (Supp. Information)

| Station # | Camera + lens details | Image size (pixels) | Resolution at center (pixels/degree) |
|---|---|---|---|
| 1 | Unknown (comercial dashboard camera) | 1600×900 | 17.28 |
| 2 | Samyang 14/2.8 + Sony a7r | 1366×768 | 9.842 |
| 3 | Samsung SDC -435 1/3", Tamron 3-8 mm, F=1,0 | 768×576 | 9.267 |
| 4 | All-sky video camera Tracer TS-506 PSC" with a 1/3" chip. | 720×576 | 3.582 (vertical direction) and 3.290 (in horizontal) |

Table 3. Mass, trajectory and radiant data for Annama's bolide. $m_{abs}$ is the absolute magnitude, $M_b$ and $M_t$ are the computed initial and terminal masses, $H_b$, $H_{max}$ and $H_e$ are the height for the beginning, maximum, and ending parts of the computed trajectory. The meteorite bulk density was measured in 3.5 g/cm$^3$ (Gritsevich et al. 2014ab). Radiant is given for equinox (2000.0). Last three columns are the velocity at infinity, geocentric and heliocentric.

| Code | $m_{abs}$ | $M_b$ (kg) | $M_t$ (kg) | $H_b$ | $H_{max}$ | $H_e$ | $\alpha_g$ (°) | $\delta_g$ (°) | $V_\infty$ (km/s) | $V_g$ | $V_h$ |
|---|---|---|---|---|---|---|---|---|---|---|---|
| FN20140419 (Kola) | 18.3 ±0.7 | 472 | 12.5 | 83.9 | 34.6 | 21.8 | 213.03±0.20 | +8.7±0.4 | 24.2±0.5 | 21.5 | 36.3 |

Table 4. Orbital elements of Annama, and Apollo asteroid 2014UR116. Equinox (2000.00).

| Code | q (AU) | a (AU) | e | i (º) | ω (º) | Ω (º) |
|---|---|---|---|---|---|---|
| FN20140419 | 0.634± 0.006 | 1.99 ± 0.12 | 0.69 ±0.02 | 14.65 ± 0.46 | 264.77 ± 0.55 | 28.611 ± 0.001 |
| *2014 UR116* | *0.563579* | *2.06962405* | *0.727689* | *6.57463* | *286.8123* | *6.0125* |

# FIGURE CAPTIONS

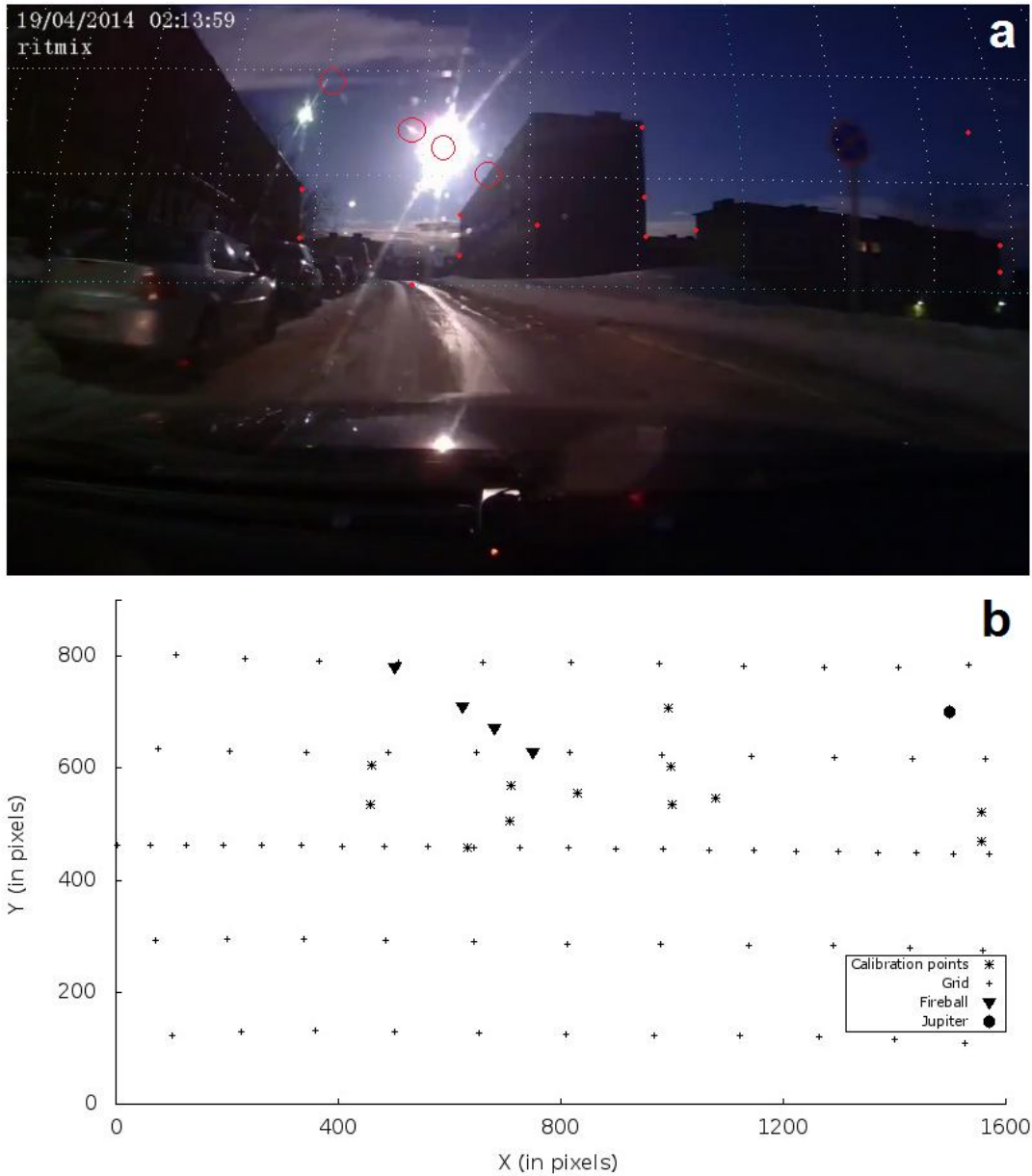

Figure 1. a) Composite image of the bolide as recorded by a dashcam in Snezhnogorsk. A calibration grid every 10º has been overlaid that contains information on the imaging geometry and camera orientation. b) The grid is shown again together with the measured points for Jupiter, the fireball and the building features that were used for trajectory calibration.

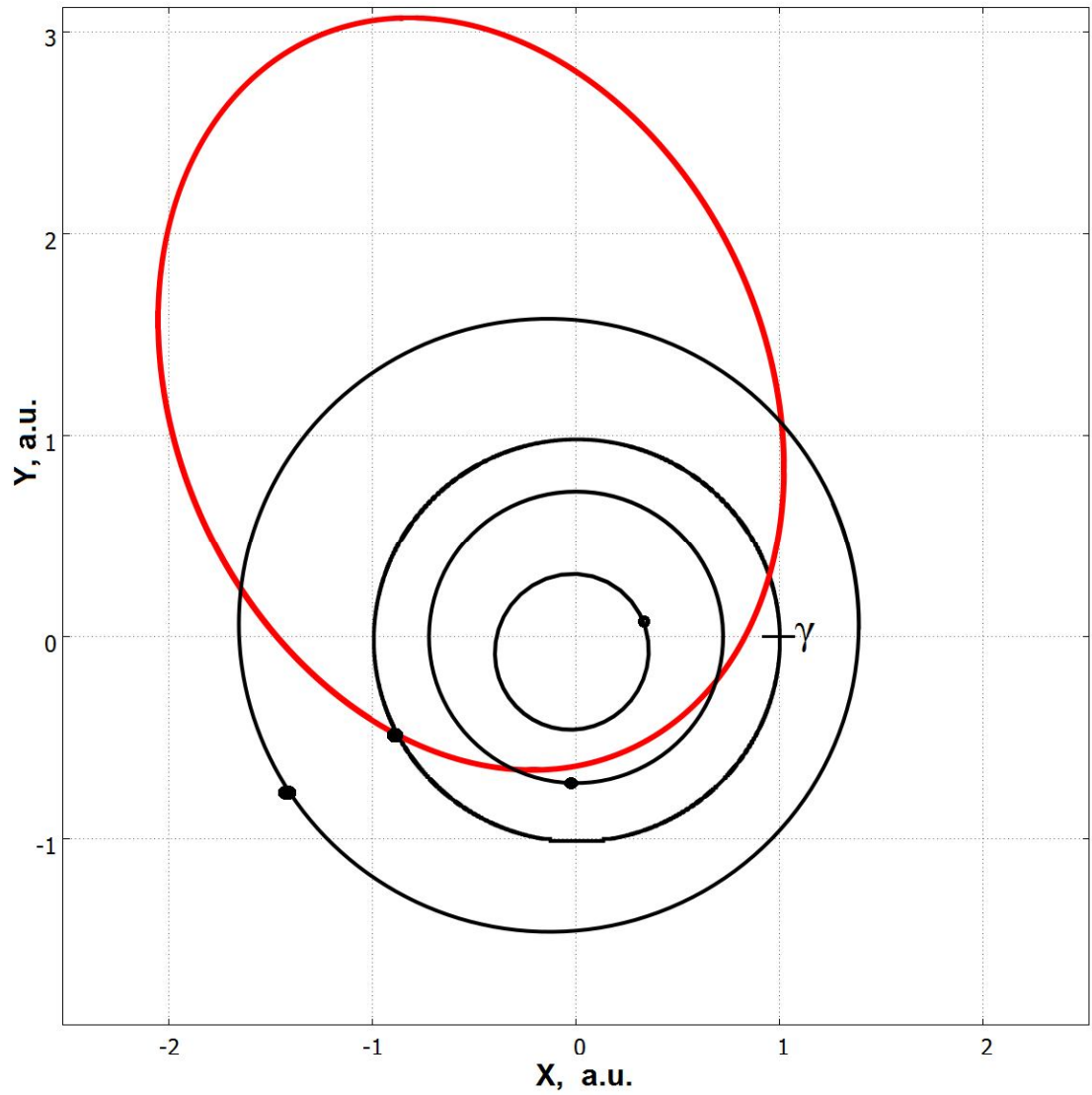

Figure 2. The heliocentric orbit of Annama meteoroid projected into the ecliptic plane and its relative position to the orbits of Mercury, Venus, Earth, and Mars. Grid corresponds to the ecliptic J2000 coordinate system.

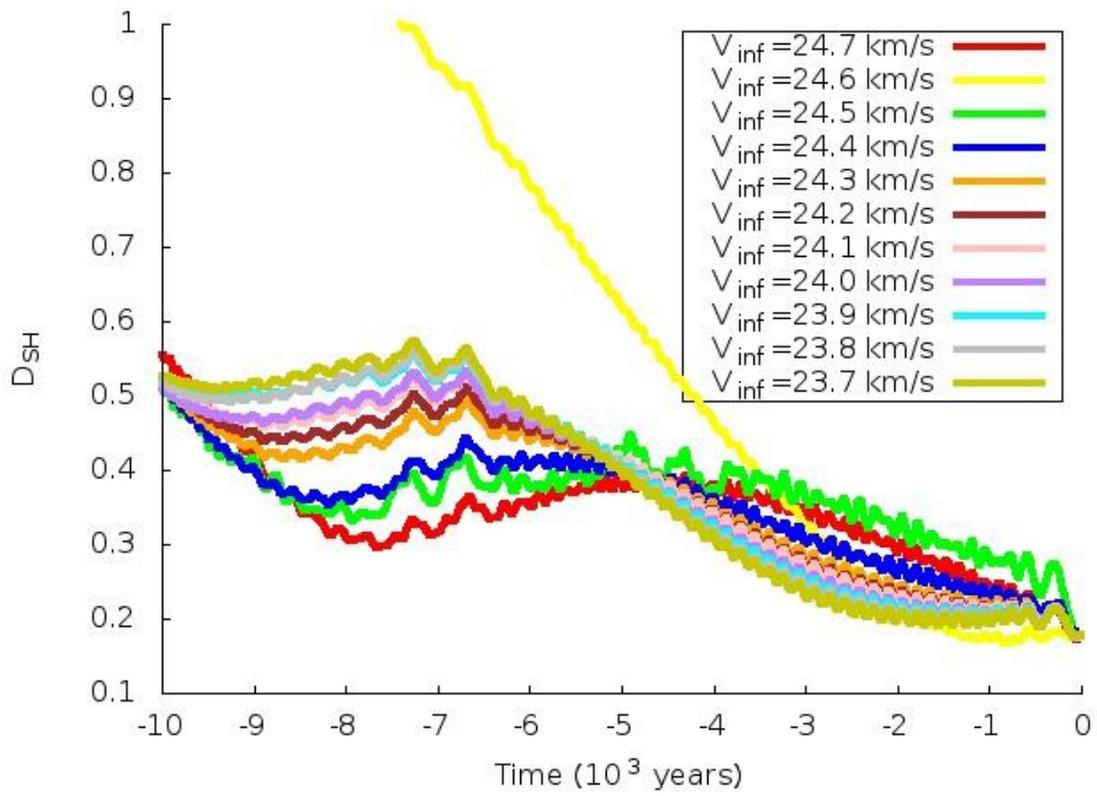

Figure 3. Southworth-Hawkins dissimilarity criterion ($D_{SH}$) comparing with different Annama's orbits generated by changing the pre-atmospheric velocity $V_{inf}$ (in km/s) in the range given by the calibration uncertainty and 2014UR116 nominal orbit over 10,000 years.

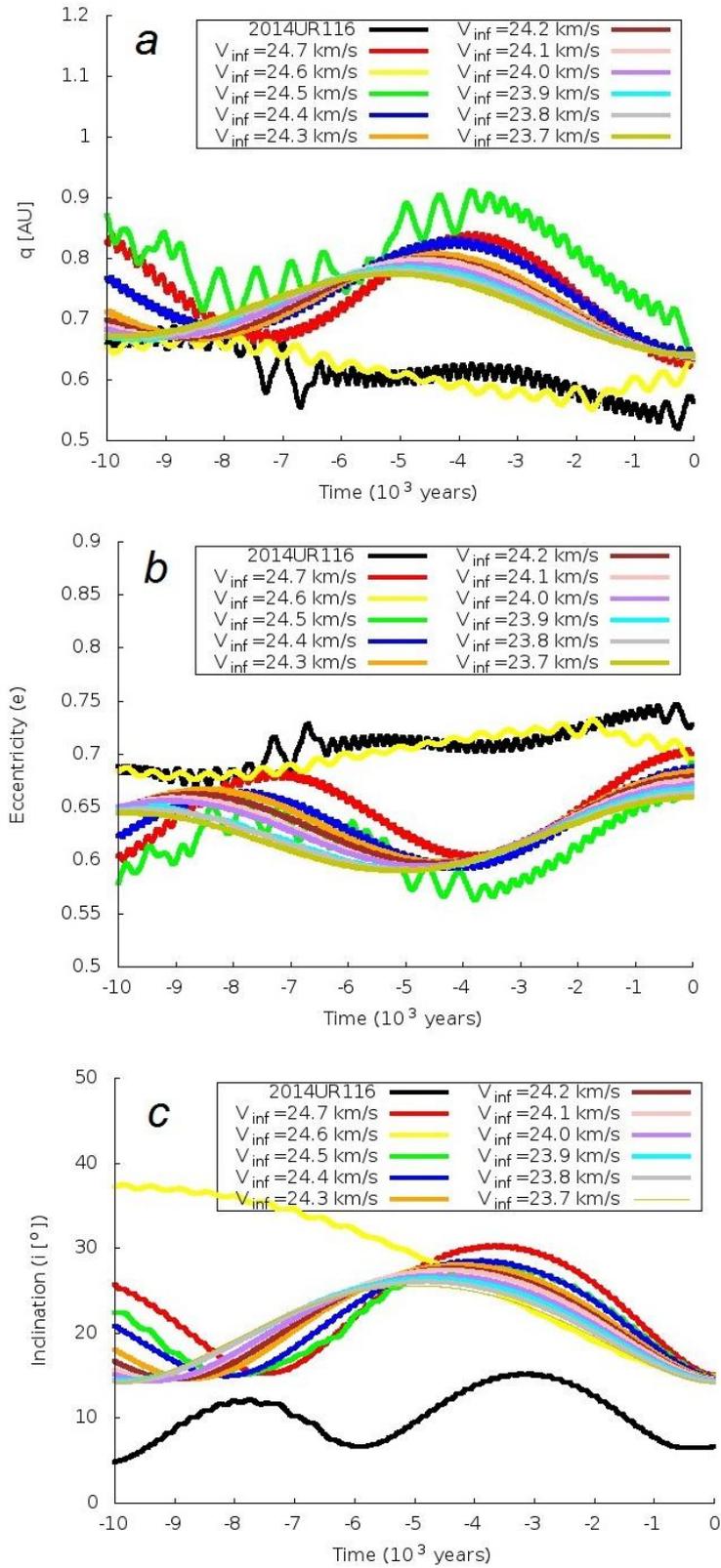

Figure 4. Numerical integration of q (graph a), eccentricity (b) and inclination (c) for several orbits of Annama obtained for the plausible pre-atmospheric velocity $V_{inf}$ (in km/s) and asteroid 2014UR116 over 10,000 years.